\documentclass[twocolumn,prl,tightenlines,superscriptaddress,showpacs]{revtex4-1}

\usepackage{amsmath}
\usepackage{amssymb,amsfonts,latexsym}
\usepackage{bm}
\usepackage[mathcal]{euscript}
\usepackage{graphicx}
\usepackage{epsfig}
\usepackage{color}
\usepackage{xfrac}
\usepackage{mathrsfs}

\begin{document}

\title{Dynamical Sub-classes of Dry Active Nematics}

\author{Li-bing Cai}
\affiliation{National Laboratory of Solid State Microstructures and Department of Physics, Nanjing University, Nanjing 210093, China}
\affiliation{Center for Soft Condensed Matter Physics and Interdisciplinary Research, Soochow University, Suzhou 215006, China}

\author{Hugues Chat\'{e}}
\affiliation{Service de Physique de l'Etat Condens\'e, CEA, CNRS Universit\'e Paris-Saclay, CEA-Saclay, 91191 Gif-sur-Yvette, France}
\affiliation{Computational Science Research Center, Beijing 100094, China}
\affiliation{Center for Soft Condensed Matter Physics and Interdisciplinary Research, Soochow University, Suzhou 215006, China}

\author{Yu-qiang Ma}
\affiliation{National Laboratory of Solid State Microstructures and Department of Physics, Nanjing University, Nanjing 210093, China}
\affiliation{Center for Soft Condensed Matter Physics and Interdisciplinary Research, Soochow University, Suzhou 215006, China}

\author{Xia-qing Shi}
\affiliation{Center for Soft Condensed Matter Physics and Interdisciplinary Research, Soochow University, Suzhou 215006, China}
\affiliation{Service de Physique de l'Etat Condens\'e, CEA, CNRS Universit\'e Paris-Saclay, CEA-Saclay, 91191 Gif-sur-Yvette, France}

\date{\today}

\begin{abstract}
We show that the dominant mode of alignment plays an important role in dry active nematics,
leading to two dynamical sub-classes 
defined by the nature of the instability of the nematic bands that characterize, in these systems, the coexistence phase
separating the isotropic and fluctuating nematic states. 
In addition to the well-known instability inducing long undulations along the band, 
another stronger instability leading to the break-up of the band in many transversal segments may arise. 
We elucidate the origin of this strong instability for a realistic model of self-propelled rods and determine the high-order nonlinear terms
responsible for it at the hydrodynamic level. 
\end{abstract}

\maketitle

Active nematics is a major topic within active matter studies.
The term loosely refers to situations where orientational nematic order typically emerges from interacting elongated self-propelled particles. 
Very different situations are in fact grouped together under this name: biological tissues
\cite{KEMKEMER,GRULER,DUCLOS,KAWAGUCHI,LADOUX}, swimming sperm cells \cite{CREPPY}, bacteria \cite{ZHOU,SHAEVITZ,DAIKI},
in vitro cytoskeleton assays \cite{SUMINO,DOGIC-NATMAT,SAGUES,SAGUES2,BAUSCH,TANIDA}, 
and man-made systems such as shaken granular rods \cite{NARAYAN}. 
As noted early on by Ramaswamy et al. \cite{SRIRAM1,SRIRAM-REV}, dry and wet active nematic systems
(where the fluid in which the particles move can or cannot be neglected) are expected to behave differently. 
Whereas the wet case is the topic of ongoing theoretical debates 
(see \cite{YEOMANS-DEFECT,MCM-DEFECT1,MCM-DEFECT2,YEOMANS-FRICTION,YEOMANS-VORTICITY,GIOMI-PRX,INTRINSIC,SRIVASTAVA,HEMINGWAY,DUNKEL,MAITRA} for recent works), 
dry active nematics are considered to be better understood.
There is in particular some consensus about the hydrodynamic description of
dilute dry active nematics, in terms of a density and a nematic tensor field:
\begin{eqnarray}
\partial_t \rho &=& D_{\rm p} \Delta \rho + D_{\rm m} ( {\bf \Gamma} : {\bf Q} )\,,
\label{hydro3-1} \\
\partial_t {\bf Q} &=& [\mu - 2\xi ({\bf Q:Q})]{\bf Q} + D_{\rm p} \Delta {\bf Q} + D_{\rm m}{\bf \Gamma} \rho
\label{hydro3-2}
\end{eqnarray}
where $\Gamma_{11}=-\Gamma_{22}\equiv \frac{1}{2}(\partial_1\partial_1-\partial_2\partial_2)$,
 $\Gamma_{12}=\Gamma_{21} \equiv \partial_1\partial_2$, ${\bf A:B}= A_{\alpha\beta} B_{\alpha\beta}$,
 $D_{\rm p}\!=\!\frac{1}{2}(D_\|\!+\!D_\perp)$, $D_{\rm m}\!=\!\frac{1}{2}(D_\|\!-\!D_\perp)$,
and $\mu, \xi$ depend on $\rho$ and particle level parameters such as the rotational, longitudinal, and tranversal 
diffusion constants $D_{\rm r}$, $D_\|$, and $D_\perp$.

\begin{figure}[t!]
\includegraphics[width=\columnwidth]{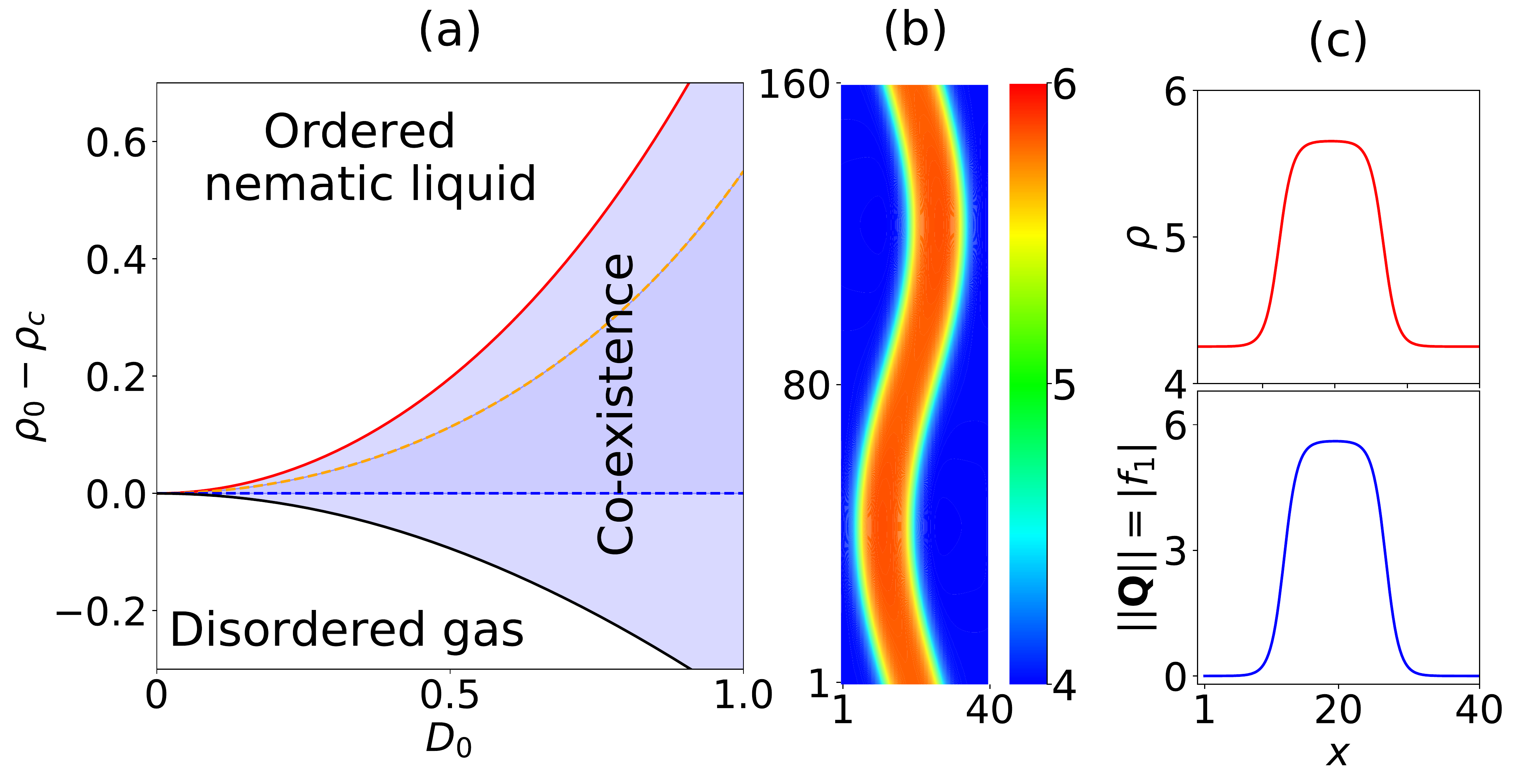}
\caption{Lowest order hydrodynamic equations (\ref{hydro3-1},\ref{hydro3-2}).
(a) Exact phase diagram in $(D_0=D_{\rm m}/D_{\rm p}, \rho_0)$ plane ($\rho_0$ is the global density) with $\mu$ and $\xi$ as derived
from kinetic equation \eqref{kinetic} \cite{SUPP} and $D_{\rm r}=2$. 
The disordered gas and the nematic liquid are separated by a coexistence phase where dense nematic bands evolve chaotically (see Movie~1 in \cite{SUPP}. This region, delimited by the two binodals (black and red solid lines), contains the two spinodals: 
The dashed blue line marks the limit of stability of the disordered gas 
(given by $\mu=0$, which, when Eqs.~(\ref{hydro3-1},\ref{hydro3-2}) are derived from \eqref{kinetic}, translates into 
$\rho=\rho_c=\frac{3\pi}{2}$ \cite{SUPP}),
while the ordered liquid is stable above the dashed yellow line. 
(b) Coexistence phase: Snapshot of a band undergoing the basic $-|{\bf u}\cdot{\bf u}'|$ instability (density field, colormap to the right). 
Parameters as in (a), with $D_0=0.75$, $\rho_0=1.01\rho_c$.
(c) Density and nematic order profiles of the band solution (parameters as in (b)).
}
\label{fig1}
\end{figure}

These equations, possibly endowed with noise terms, are known to describe correctly 
the main qualitative features of dry and dilute active nematics in 2D (two space dimensions, 
where most work has been performed) \cite{NEMAMESO,NEMANEW,BASKARAN,SCM-NJP}: 
the homogeneous nematic liquid with quasi-long range order and giant number fluctuations present at large density and weak noise is separated from 
the disordered gas by a coexistence phase characterized by the spatiotemporally chaotic dynamics of high-density high-order nematic bands
(Fig.~\ref{fig1}a and Movie~1 in \cite{SUPP}).
Most of these phenomena have been observed recently in experimental systems \cite{DAIKI,BAUSCH,TANIDA}.

As a matter of fact, Eqs.~(\ref{hydro3-1},\ref{hydro3-2}) were derived from models having rather different
interactions between particles. In the Vicsek-style model of \cite{NEMA1,NEMANEW}, alignment is explicit  
and appears as the rotation of velocities upon collision (``rotational alignment"). 
The kinetic theory of \cite{SCM-NJP} is an active version of the 
Doi-Onsager theory of rods, with alignment resulting from avoiding overlaps between elongated objects (``positional alignment"). 

Here we show that the dominant mode of alignment (rotational or positional) actually plays an important role in the collective dynamics of 2D dry active nematics.
In particular, the spatiotemporal chaos characterizing the coexistence phase emerges from two different instabilities 
of the nematic band solution 
for rotational and positional alignment. We show that these two dynamical sub-classes can be observed 
within a generic self-propelled rods model when varying their velocity reversal rate, and in Vicsek-style models with tailored alignment modes.
The two classes can be accounted for at the kinetic level, but not at the standard hydrodynamic level of Eqs.~(\ref{hydro3-1},\ref{hydro3-2}). 
Higher-order hydrodynamic descriptions, on the other hand, display the two types of behavior
and we determine the nonlinear terms determining to which class a given system belongs. 
We finally discuss the experimental relevance of our findings and their possible consequences for asymptotic correlations and fluctuations. 

\begin{figure}[t!]
\includegraphics[width=\columnwidth]{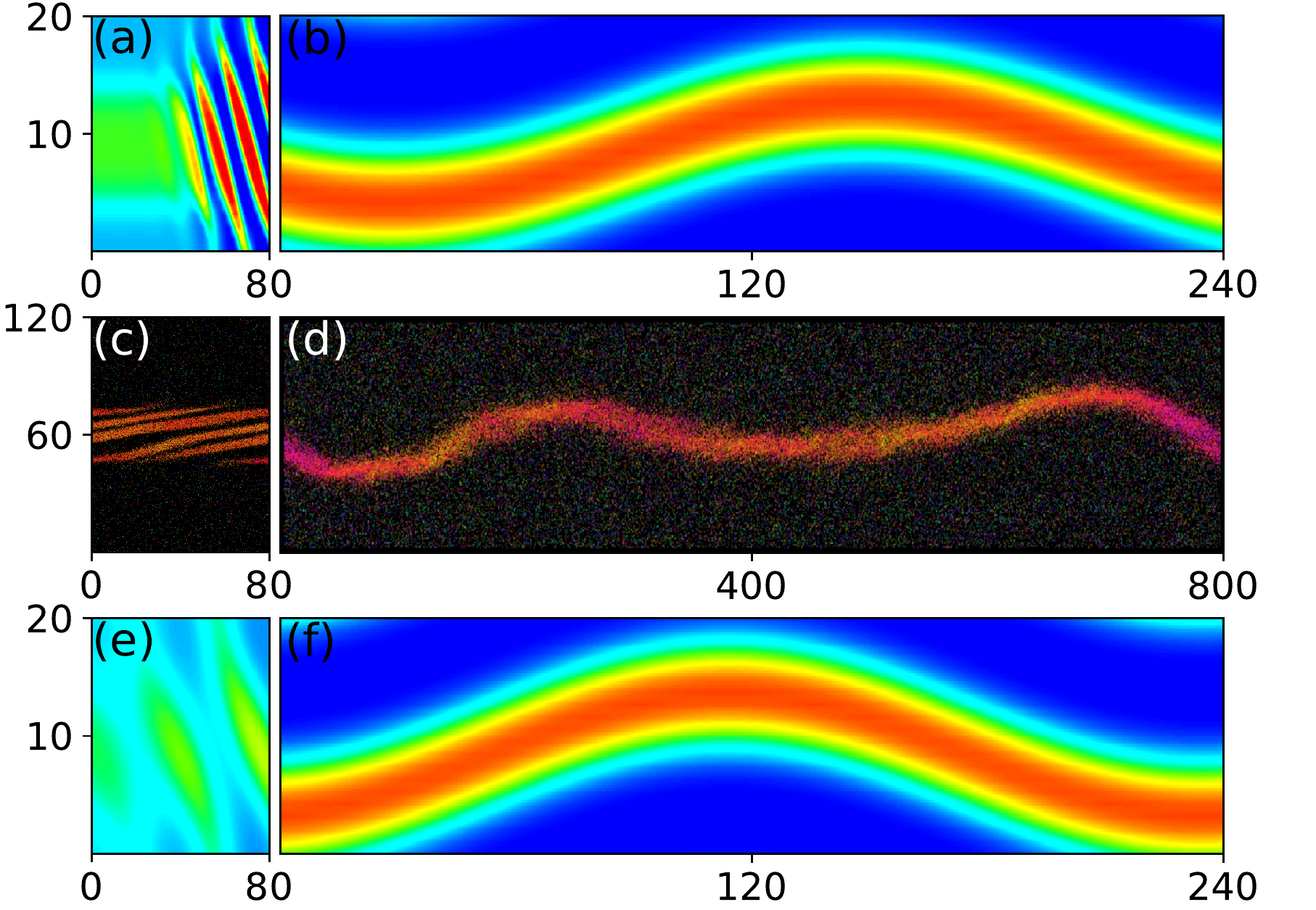}
\caption{Snapshots at the early stage of the band instability. Left panels: transversal breakup into small fragments characteristic
of the $|{\bf u}\times{\bf u}'|$ instability. Right panels: default long-wavelength $-|{\bf u}\cdot{\bf u}'|$ instability.
(a,b): Kinetic equation \eqref{kinetic} truncated at 10 Fourier modes (no significant difference was detected when using more modes) with 
$|{\bf u}\times{\bf u}'|$ potential \eqref{uxu-pot} (a) and $-|{\bf u}\cdot{\bf u}'|$ potential \eqref{u.u-pot} (b)
 ($D_{\rm p}=2.4$, $D_{\rm m}=1.2$, $D_{\rm r}=2$, $\rho_0=1.01\rho_c$).
 (c,d): Continuous-time Vicsek-model (\ref{position},\ref{angle}) with potential 
 $U(\theta,\theta')=|{\bf u}(\theta)\times{\bf u}(\theta')|$ (c) and $U(\theta,\theta')=-|{\bf u}(\theta)\cdot{\bf u}(\theta')|$ (d)
 ($K=1$, $\sigma=0.45$, $\rho_0=1.5$, $v_0=5$).
 (e,f): Hydrodynamic equations (\ref{hydro5-1}-\ref{hydro5-3}) with parameters as derived from kinetic equation \eqref{kinetic} with
 potential \eqref{uxu-pot} (a) and potential \eqref{u.u-pot} (b) (basic parameters as in (a,b)).
 Data shown: density field, colormap as in Fig.~\ref{fig1} for panels (a,b,e,f); in (c,d) particles are colored by their orientation, as in Fig.~\ref{fig4}.
}
\label{fig2}
\end{figure}

At the kinetic (Fokker-Planck) level, rotational and positional alignment are distinguished by the ``self-consistent interaction potential"
$w(\textbf{r},\textbf{u})$
entering the equation governing $f({\bf r},{\bf u},t)$, the probability of finding particles at location ${\bf r}$, 
with director $\pm{\bf u}$, at time $t$ \cite{SCM-NJP}:
\begin{eqnarray}
\label{kinetic}
\partial_t f(\textbf{r},\textbf{u},t) &=& \nabla[D_{\|}{\bf u}{\bf u}\nabla +D_\perp ({\bf I} - {\bf u}{\bf u})\nabla] f({\bf r},{\bf u})  \nonumber \\
&+& \mathscr{R}[D_{\rm r}\mathscr{R}f({\bf r},{\bf u})+D_{\rm r}f({\bf r},{\bf u})\mathscr{R}w({\bf r},{\bf u})]
\end{eqnarray}
where $\mathscr{R} ={\bf u}\times\partial_{\bf u}$ is the rotation operator. 
For positional alignment (Doi-Onsager theory), one has 
\begin{equation} 
\label{uxu-pot}
w(\textbf{r},\textbf{u})=l^2\int{\rm d}{\bf u}'|{\bf u}\times{\bf u}'|f({\bf r},{\bf u}')
\end{equation}
while for rotational alignment 
\begin{equation} 
\label{u.u-pot}
w(\textbf{r},\textbf{u})=-l^2\int{\rm d}{\bf u}'|{\bf u}\cdot{\bf u}'|f({\bf r},{\bf u}') \;.
\end{equation}
($l$ is the particle's length or interaction range.)
The phase diagram of Eq.~\eqref{kinetic} is similar to that of Eqs.(\ref{hydro3-1},\ref{hydro3-2}) shown in Fig.~\ref{fig1}a \cite{SCM-NJP}.
In particular, Eq.~\eqref{kinetic} also possesses a nematic band solution that is always unstable leading to a chaotic 
coexistence phase. 
Strikingly, the instability of the band differs strongly depending on the interaction potential $w$. 
For the $-|{\bf u}\cdot{\bf u}'|$ rotational potential the band slowly develops long wavelength longitudinal modulations 
(Fig.~\ref{fig2}b, Movie~2 in \cite{SUPP}),
similar to that shown in Fig.~\ref{fig1}b for the simple hydrodynamic equations (\ref{hydro3-1},\ref{hydro3-2}). 
For the $|{\bf u}\times{\bf u}'|$ positional potential, on the other hand, the band breaks rather suddenly into a multitude of small 
transverse segments (Fig.~\ref{fig2}a, Movie~3 in \cite{SUPP}), as reported already in \cite{SCM-NJP}.
In fact, in this last case, the band solution is unstable even in one-dimensional domains \cite{1D}. 
The two different band instabilities observed for positional ($|{\bf u}\times{\bf u}'|$) and rotational ($-|{\bf u}\cdot{\bf u}'|$) alignment 
change the spatiotemporally chaotic dynamics of the coexistence phase and thus define two ``dynamical sub-classes'' of dry active nematics.

The same phenomenology is observed for simple continuous-time Vicsek-style models where point particles
with position ${\bf r}_i$ and orientation $\theta_i$ move at constant speed $v_0$, and align nematically with neighbors:
\begin{eqnarray}
\dot{\bf r}_i &=& \pm v_0 {\bf u}(\theta_i) \label{position} \\
\dot{\theta}_i &=& -K \sum_{j\sim i} \partial_{\theta_i} U(\theta_i,\theta_j) + \sigma \xi_i \;. \label{angle}
\end{eqnarray}
In \eqref{position}, $\pm$ codes for velocity reversals at some finite rate $r$ and ${\bf u}(\theta_i)$ is the unit vector along $\theta_i$.
In \eqref{angle}, $K$ is a coupling constant, $\xi_i$ is a uniform white noise in $[-1,1]$, $\sigma$ is the noise strength, 
and $U$ is the interaction potential, which can take the form
$U(\theta,\theta') =  |{\bf u}(\theta)\times{\bf u}(\theta')|$ or $U(\theta,\theta') =  -|{\bf u}(\theta)\cdot{\bf u}(\theta')|$. 
In both cases, this model exhibits the usual phenomenology of dry dilute active nematics, 
with a coexistence phase made of chaotically evolving nematic bands. 
But here again, the band breaks in transversal segments for the $|{\bf u}\times{\bf u}'|$ potential, whereas it displays a long-wavelength instability 
for the $-|{\bf u}\cdot{\bf u}'|$ interaction (Fig.~\ref{fig2}c,d, Movies~4,5 in \cite{SUPP}).
Thus the existence of two sub-classes of dry active nematics 
is {\it not} due to the approximations used to describe these systems by simple kinetic equations such as \eqref{kinetic}.
The two instability modes are rooted in the microscopic fluctuating level. 

We now turn to the hydrodynamic equations that can be derived from kinetic equations such as \eqref{kinetic}. 
At the simplest non-trivial order usually considered, 
whether Fokker-Planck or Boltzmann kinetic equations are used,
and irrespective of the interaction potential considered, one always finds Eqs.~(\ref{hydro3-1},\ref{hydro3-2})
albeit with different expressions of the transport coefficients. 
The nematic band solution of  Eqs.~(\ref{hydro3-1},\ref{hydro3-2}) is known in closed form \cite{VICSEK-RODS-HYDRO,NEMAMESO} 
and was shown in \cite{NEMANEW,PERUANI} 
to be always unstable, in a large-enough system, to a long-wavelength instability of the $-|{\bf u}\cdot{\bf u}'|$ type described above.
Moreover, varying systematically the parameters,
we found that the band solution
 is always ``1D stable" \cite{1D} where it exists (not shown). 
Thus $|{\bf u}\times{\bf u}'|$ type instability is absent and the lowest-order hydrodynamic level of 
dry active nematics is unable to account for the two-dynamical sub-classes.

\begin{figure}[t!]
\includegraphics[width=\columnwidth]{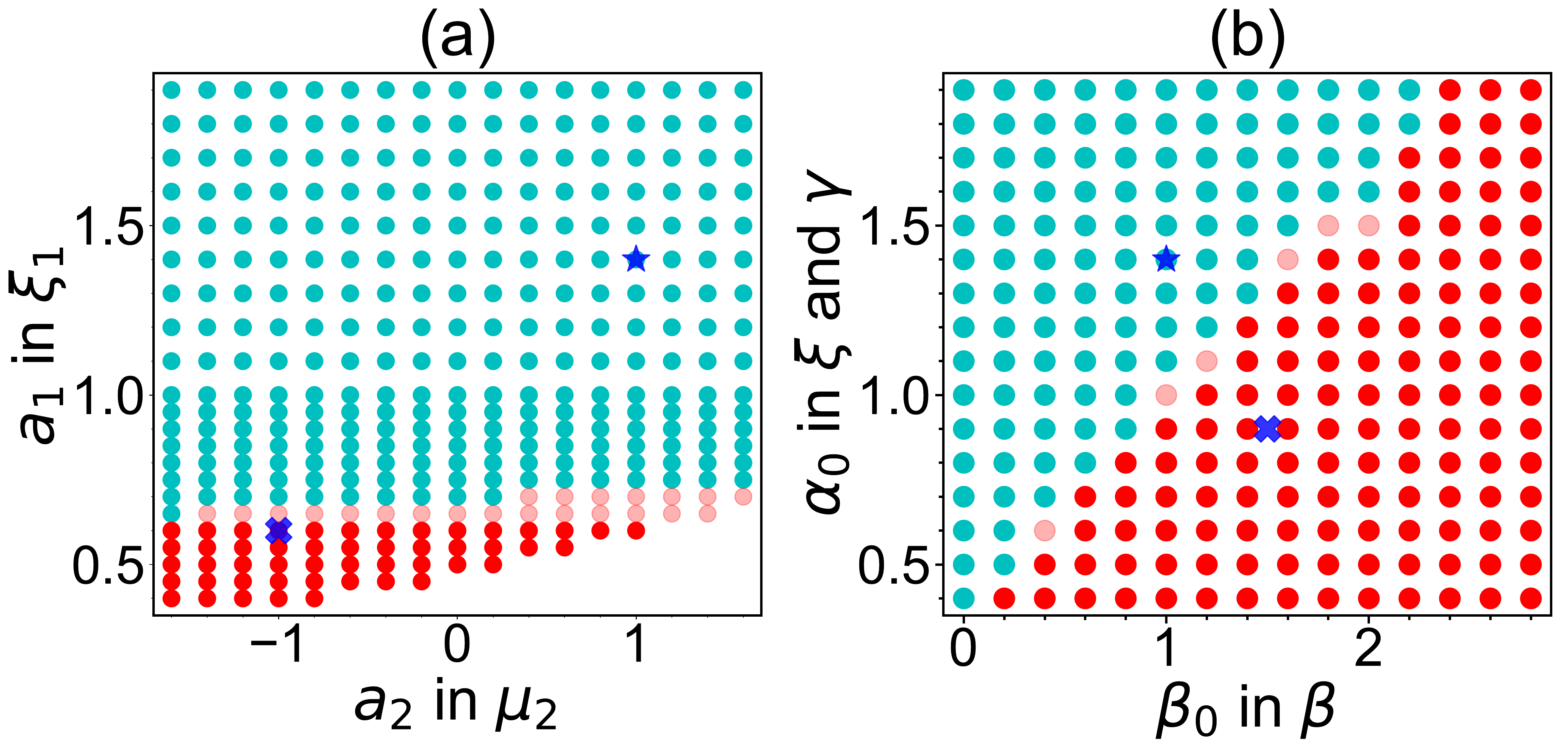}
	\caption{
		(a) 1D stability of the band solution of Eqs.~(\ref{hydro5-1}-\ref{hydro5-3}) in the $(\mu_2,\xi_1)$  plane.
		Writing $\xi_1=a_1\frac{8}{3\pi}D_{\rm r}$ and $\mu_2=-16(1+\frac{2a_2}{15\pi}\rho)D_{\rm r}$, we actually vary $a_1$ and $a_2$.
Other parameters: $D_{\rm m} = 0.25$, $D_{\rm p}=0.5$, $D_{\rm r}=4$, and $a_3=1$, global density $\rho_0 = 1.01\rho_c$, 
with $\rho_c = {3\pi}/{2}$ (see \cite{SUPP}). Red dots: 1D instability, the $|{\bf u}\times{\bf u}'|$ instability dominates. Cyan dots: 1D stability, only the $|{\bf u}\cdot{\bf u}'|$ instability is present. (Pink dots indicate a 1D stable (but 2D unstable) band solution with the nematic order no longer along the band, and the empty region denotes unphysically high order.) 
	(b) same as (a) but for Eq.~\eqref{hydro5-4} in the $(\alpha_0,\beta_0)$ plane where 
	$\beta=\beta_0D_{\rm m}\chi/2$, $\xi=\alpha_08D_{\rm r}\chi/3\pi$ and $\gamma=\alpha_0D_{\rm m}\chi/4$, with $\chi = 5/(15\pi+2\rho_0)$. 
	The ``$\times$" and ``star" symbols correspond to the parameters as derived from kinetic equation (3) using potentials (4) and (5), respectively.
		These results are not sensitive to the system size.
	}
\label{fig3}
\end{figure}

We now discuss hydrodynamic theories at some higher-order than Eqs.~(\ref{hydro3-1},\ref{hydro3-2}). A clean way to proceed 
is to use a Ginzburg-Landau scaling ansatz to truncate and close the hierarchy of equations obtained 
when expressing the kinetic equation in terms of 
Fourier modes $f_k(\mathbf{r}) = \int_{-\pi}^\pi d\theta e^{2 {\rm i} k \theta} f(\mathbf{r},{\bf u}(\theta))$ \cite{BGL}. 
(In the diffusive limit of frequent velocity-reversals considered here, the odd modes vanish.) 
The scaling ansatz is $f_{k} \sim \epsilon^{|k|}$,
$\partial_t \sim \nabla^2 \sim (\delta\rho)^2 \sim \epsilon^2$, where $\epsilon$ is a small parameter characterizing
the distance to the onset of order, i.e. the stability limit of the disordered solution. 
Eqs.~(\ref{hydro3-1},\ref{hydro3-2}) are the result of this procedure applied at the first non-trivial order $\epsilon^3$. 
The density field $\rho = f_0$, and the nematic field ${\bf Q}$ is equivalent to $f_1$ via $Q_{xx}=\frac{1}{2}\Re\{f_1\}$, $Q_{xy}=\frac{1}{2}\Im\{f_1\}$.
At the next order $\epsilon^4$, one obtains the following closed equations \cite{SUPP}
\begin{eqnarray}
\!\!\!\! \partial_t \rho &=& D_{\rm p} \triangle\rho + D_{\rm m} \Re\{\triangledown^{*2}f_1\} \label{hydro5-1}\\  
\!\!\!\! \partial_t f_1 &=& \mu f_1 \!-\! \xi_1 f_1^*f_2 \!+\! \tfrac{1}{2} D_{\rm m} (\triangledown^2\rho \!+\! \triangledown^{*2}f_2) 
\!+\! D_{\rm p}\triangle f_1 \label{hydro5-2} \\ 
\!\!\!\! \partial_t f_2 &=&  [\mu_2 \!-\! \xi_2 |f_1|^2] f_2 \!+\! \tfrac{1}{2}D_{\rm m}\triangledown^2f_1 \!+\! \beta_2 f_1^2  \!+\! D_{\rm p} \triangle f_2  \label{hydro5-3}
\end{eqnarray}
where we have kept, for convenience, the complex fields $f_1$ and $f_2$, the $^*$ superscript denotes complex conjugates, 
and the complex gradient $\triangledown=\partial_x + i \partial_y$, so that the Laplacian is $\triangle=\triangledown\triangledown^*$.
The form of these equations does {\it not} depend on the particular potential $w$ considered. 
Their phase diagram in the basic $(\rho_0,D_0)$ parameter plane remains similar to Fig.~\ref{fig1}a. 
Actually, only a few of their transport coefficients, namely $\xi_1$, $\mu_2$, and $\xi_2$, depend on $w$ \cite{SUPP}.
Remarkably, simulations performed with the coefficients derived from the positional and the rotational potential show that
these equations then account for the two corresponding instability modes of the band solution (Fig.~\ref{fig2}e,f and Movies~6,7 in \cite{SUPP}). 
Since $\xi_2$ does not change much with $w$ and the term associated with this coefficient is anyway of relatively high order, 
$\mu_2f_2$ and $\xi_1 f_1^*f_2$ must be the terms determining the relevant dynamical sub-class. 
We confirmed this by studying the 1D stability of the band solution in the $(\mu_2,\xi_1)$ plane.
A large region where the band is 1D-unstable and the $|{\bf u}\times{\bf u}'|$ positional instability dominates is found (Fig.~\ref{fig3}a).

To gain a better understanding of this,
it is convenient to reduce \eqref{hydro5-2} and \eqref{hydro5-3} to a single equation by enslaving $f_2$ to $f_1$ 
($\mu_2f_2=-\frac{1}{2}D_{\rm m}\triangledown^2f_1\!-\!\beta_2 f_1^2$ at order $\epsilon^4$). Eq.~\eqref{hydro5-2} then becomes:
\begin{eqnarray}
\partial_t f_1 &=& [\mu-\xi |f_1|^2] f_1 +\tfrac{1}{2}D_{\rm m} \triangledown^2\rho + D_{\rm p}\triangle f_1 \nonumber \\
&&+\beta\triangledown^{*2}f_1^2 -\gamma f_1^*\triangledown^2 f_1
\label{hydro5-4}
\end{eqnarray}
which is identical to \eqref{hydro3-2} except for the last two higher-order terms. 
Varying the coefficients of these terms, we confirm that they decide, together with the $\xi|f_1|^2]f_1$ term, 
the dominating band instability (Fig.~\ref{fig3}b).

\begin{figure}[t!]
\includegraphics[width=\columnwidth]{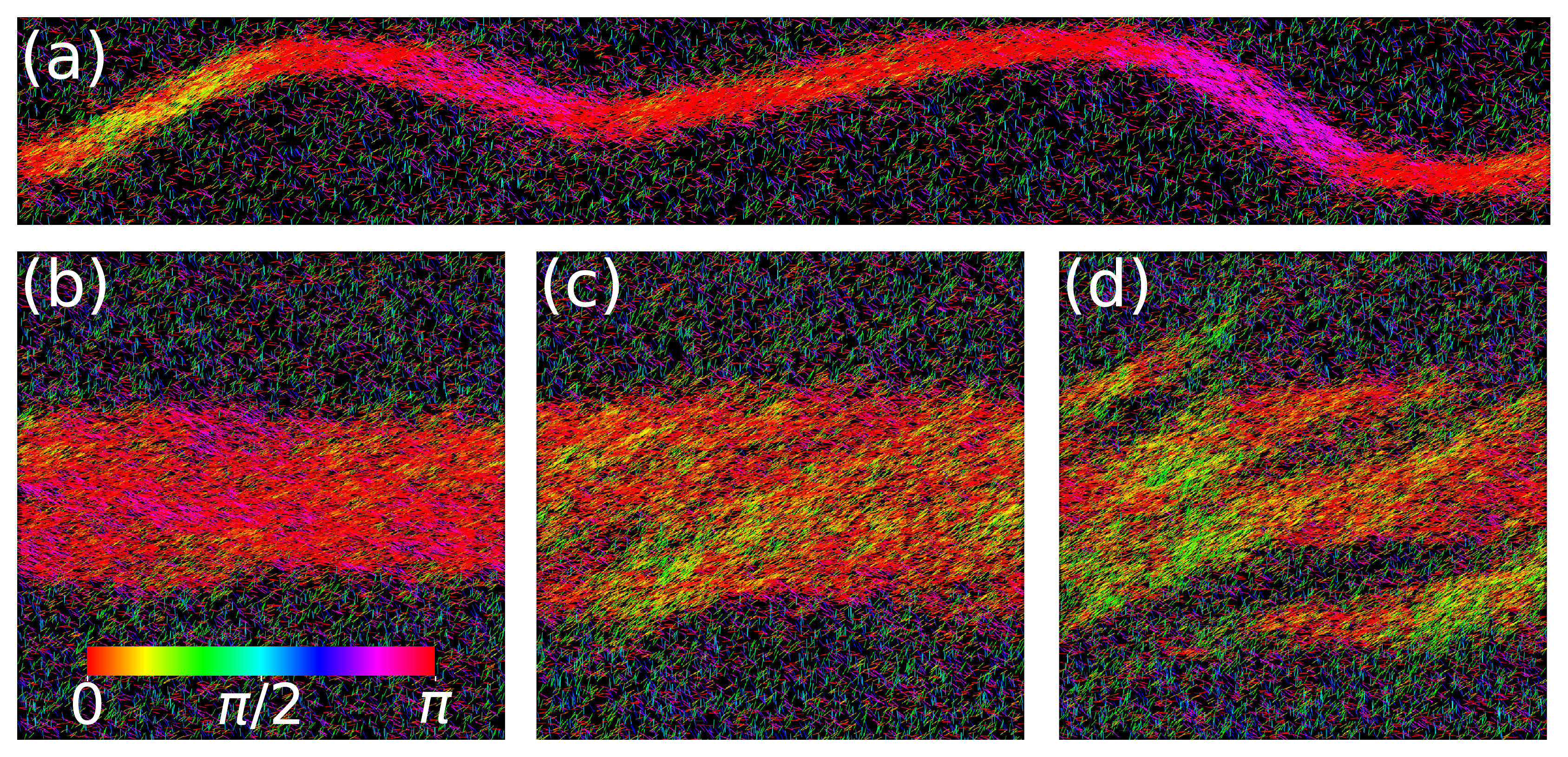}
\caption{Instability of the nematic band solution in a generic model of self-propelled rods.
(a): at low reversal rate (here $r=0.015$), only the long-wavelength $-|{\bf u}\cdot{\bf u}'|$ is observed (long system of size $2845\times 355$, packing fraction 0.5).
(b-d): at large-enough reversal rate (here $r=0.075$) the $|{\bf u}\times{\bf u}'|$ instability breaks the bands in multiple transversal segments (thick band in a square domain of linear size 920, packing fraction 0.6). The 3 snapshots show the growth of the instability.
Other parameters (defined in \cite{SUPP}): $N=32000$ rods of length 16 ($a=15$), $S=25$, $\mu_\|/\mu_\perp=1$, $\eta=0.1$. 
Rods are colored by their orientation, following the colormap in (b).
}
\label{fig4}
\end{figure}

We now come back to the microscopic, fluctuating level. The two types of alignment (positional and rotational) can be seen as limit cases. 
In realistic active nematics such as self-propelled elongated objects interacting by volume exclusion, the two alignment modes are probably 
present with different weights depending on details of the dynamics. We studied a generic model of 
overdamped self-propelled rods introduced in \cite{RODS} to which we added velocity reversals. 
Using parameter values shown in \cite{RODS} to yield stable
nematic bands at moderate system size without velocity reversals, we studied the effect of the reversal rate $r$. 
For $r$ small, the band, simulated in a large-enough system, shows the long modulations typical of the
$-|{\bf u}\cdot{\bf u}'|$ instability (Fig.~\ref{fig4}a, Movie~8 in \cite{SUPP}). For fast-enough reversals, on the other hand, the band breaks into many transversal 
short pieces, the signature of the $|{\bf u}\times{\bf u}'|$ instability (Fig.~\ref{fig4}b-d, Movie~9 in \cite{SUPP}). 

This finding can be rationalized by studying
interactions between rods (details about the following arguments can be found in \cite{SUPP}). Interactions are of two types:
those induced by diffusion that, for the model considered here, as in the Doi-Onsager theory of rods,
yield an effective interaction potential
\begin{equation} 
\label{rods-uxu-pot}
w_c(\textbf{r},\textbf{u})\propto F l^3\int{\rm d}{\bf u}'|{\bf u}\times{\bf u}'| f({\bf r},{\bf u}')
\end{equation}
where $l$ is the length of rods and $F$ is the typical magnitude of the force when they overlap.
The other interactions are generated by the self-propulsion along the rods' axes. 
Upon collision onto another, a rod rotates. The resulting collision-induced rotation rate can be written
\begin{equation} 
\label{rods-coll-rate}
\mu_r   \frac{Fl^2v_0\tau}{8} \int{\rm d}{\bf u}' ({\bf u}\cdot{\bf u}')({\bf u}\times{\bf u}') f({\bf r},{\bf u}')
\end{equation}
where $\mu_r$ is the effective rotational mobility, $v_0$ the self-propulsion force, and $\tau$ the typical collision time. 
When the reversal rate $r\gtrsim v_0/l$, $\tau$ is mainly controlled by the reversal time. 
The effective angular potential induced by these rotations can then be written as
\begin{equation} 
\label{rods-coll-pot}
w_r(\textbf{r},\textbf{u})\propto -F l^2 \frac{v_0}{r} \int{\rm d}{\bf u}' \cos [2(\theta-\theta')] f({\bf r},{\bf u}')
\end{equation}
Thus, the kinetic equation for the active rods studied here could be identical to \eqref{kinetic} 
but with the potential $w$ replaced by some linear combination of $w_c$ and $w_r$ \cite{NOTE}. 
Potential \eqref{rods-coll-pot} is not strictly identical to the simple one \eqref{u.u-pot} used earlier,
but it has a similar shape and influence on the band stability.
We checked at kinetic and hydrodynamic levels 
that taking $w=w_c+w_r$ and varying the reversal rate $r$ accounts for the observations 
reported in Fig.~\ref{fig4}: for large $r$, $w_c$ dominates and the 
$|{\bf u}\times{\bf u}'|$-type positional instability is present, whereas at small $r$ only the long-wavelength $|{\bf u}\cdot{\bf u}'|$-type instability
remains (not shown).
We stress that the factor deciding which instability dominates is {\it not} the reversal rate $r$ per se, 
but rather the fact that $r$ regulates the relative weight of the two alignment modes in our rods model. 
This conclusion is reinforced by our observation that, in all the other systems studied 
here, once the interaction potential is chosen, results are unchanged if one goes away from the fast-reversals limit and studies finite and even zero reversal rates \cite{TBP}.

To summarize, we have shown the existence of two dynamical sub-classes for dry and dilute active nematics, 
which are best characterized by the instability mode of the nematic band solution present in the coexistence phase of such systems. 
Whereas the well-known instability inducing long undulations along the band 
is always present, another, stronger instability leading to the break-up of the band in many transversal segments may arise. 
Our results, obtained at microscopic, kinetic, and hydrodynamic levels on a variety of systems, are robust \cite{NOTE2}. 
In particular, they do not depend on our choice to treat here Fokker-Planck kinetic equations: we have obtained similar results
with Boltzmann equations \cite{TBP}.

The well-known simple deterministic hydrodynamic theory of Eqs.~(\ref{hydro3-1},\ref{hydro3-2})
cannot account for the strong instability and is thus deceptively universal. 
However, higher-order theories can exhibit both band instabilities and we have elucidated the nonlinear terms
deciding which instability is dominant.
At the level of elementary mechanisms at play in realistic models or experiments, one can expect both positional and rotational alignment
to be present albeit with varying relative weight. This weight, and thus the sub-class to which a given system belongs, could be gauged by
considering the outcome of binary interactions, such as performed for motility assays in \cite{SUMINO,BAUSCH,TANIDA}: 
if alignment is relatively weaker for large angles between particles, an effective 
$|{\bf u}\times{\bf u}'|$ ``positional" potential is probably at play. If alignment is strong at large angles, then the effective  
potential is probably of the ``rotational" $-|{\bf u}\cdot{\bf u}'|$ type. For instance, in the actomyosin assay of \cite{BAUSCH},
the binary collision statistics between filaments maybe interpreted as being of the $-|{\bf u}\cdot{\bf u}'|$ type, 
prefiguring long-wavelength undulations of nematic bands.

Whether the two dynamical sub-classes of dry active nematics constitute two bona fide universality classes 
ultimately depends on whether correlations and fluctuations are qualitatively different in the associated asymptotic phases. 
In both cases, the band instability leads to a chaotic coexistence phase. 
Even though these chaotic regimes look different to the eye (compare Movies~10 and 11 in \cite{SUPP}),
we have so far been unable to find {\it qualitative} differences in their correlation functions and spectra.
Similarly, one could investigate whether the dominant band instability has some influence on the scaling laws
and anomalous fluctuations characterizing the homogeneous nematic fluid phase \cite{SRIRAM2,SRIRAM3}.
These difficult questions are the subject on ongoing work.

\acknowledgments
We thank B. Mahault, A. Patelli, and C. Nardini for a critical reading of this manuscript. 
This work is partially supported by ANR project Bactterns, FRM project Neisseria, and the National Natural Science Foundation of China (grant No. 11635002 to X.-q.S. and H.C., grants No. 11474210 and No. 11674236 to X.-q.S., Grants No. 11474155 and No. 11774147 to Y.-q.M.).


\begin{thebibliography}{99}
\bibitem{NOTE} The simple calculations presented in \cite{SUPP} do not constitute a quantitative treatment of what is a difficult problem. 
Nevertheless, we believe they capture the key qualitative factors governing the structure of the effective potential and how it varies with reversal rate.
\end{thebibliography}
\end{document}